\documentclass[sn-mathphys,Numbered]{sn-jnl}

\usepackage[T1]{fontenc}
\usepackage[utf8]{inputenc} 
\usepackage{graphicx}%
\usepackage{multirow}%
\usepackage{amsmath,amssymb,amsfonts}%
\usepackage{amsthm}%
\usepackage{mathrsfs}%
\usepackage[title]{appendix}%
\usepackage{xcolor}%
\usepackage{textcomp}%
\usepackage{manyfoot}%
\usepackage{booktabs}%
\usepackage{algorithm}%
\usepackage{algorithmicx}%
\usepackage{algpseudocode}%
\usepackage{listings}%

\theoremstyle{thmstyleone}%
\theoremstyle{thmstyletwo}%

\theoremstyle{thmstylethree}%

\begin{document}

\title[Beyond Classical Models: Statistical Physics Tools for the Analysis of Time Series in Modern Air Transport]{Beyond Classical Models: Statistical Physics Tools for the Analysis of Time Series in Modern Air Transport}

\author[a]{\fnm{Felipe Olivares} \sur{}}
\email{olivaresfe@gmail.com}
\author[b]{\fnm{Massimiliano Zanin} \sur{}}
\email{massimiliano.zanin@gmail.com}

\affil[a]{\orgdiv{Instituto de F\'isica Interdisciplinar y Sistemas Complejos IFISC (CSIC-UIB)}, 
\orgname{Campus UIB}. 
\city{Palma} 
\postcode{07122}. 
\country{Spain}}

\affil[b]{\orgdiv{Instituto de F\'isica Interdisciplinar y Sistemas Complejos IFISC (CSIC-UIB)}, 
\orgname{Parc Bit}. 
\city{Palma} 
\postcode{07120}. 
\country{Spain}}

\abstract{Within the continuous endeavour of improving the efficiency and resilience of air transport, the trend of using concepts and metrics from statistical physics has recently gained momentum. This scientific discipline, which integrates elements from physics and statistics, aims at extracting knowledge about the microscale rules governing a (potentially complex) system when only its macroscale is observable. Translated to air transport, this entails extracting information about how individual operations are managed, by only studying coarse-grained information, e.g. average delays. We here review some fundamental concepts of statistical physics, and explore how these have been applied to the analysis of time series representing different aspects of the air transport system. In order to overcome the abstractness and complexity of some of these concepts, intuitive definitions and explanations are provided whenever possible. We further conclude by discussing the main obstacles towards a more widespread adoption of statistical physics in air transport, and sketch topics that we believe may be relevant in the future.}

\maketitle
\tableofcontents
\section{Introduction}\label{intro}

Modern air transport is among the most complex engineered systems on Earth, with thousands of flights, passengers, crews, and controllers interacting across congested airports and crowded skies on a daily basis \cite{janic2000air, schmitt2016air}. One of its central challenges, and the focus of Air Traffic Management (ATM), is the safe and efficient use of the limited capacity of airports and airspaces \cite{cook2007european, arblaster2018air}. There is only so much traffic that these systems can accommodate within a given time frame; when this limit is reached, even a small disruption, like a plane arriving later than expected, can set off a chain reaction. This can delay other flights and cause a cascade of further disruptions, akin to a domino effect, triggering several hours with accumulated delays and large economic costs~\cite{baik2010estimation, peterson2013economic}. 
Even worse, delays and unexpected congestion can contribute to safety-related events, through for instance an increased workload of air traffic controllers, making these aspects intermingled \cite{gifford1991airport, dy2024airspace}.

Researchers have traditionally tried to understand the behaviour of the system using models and large-scale simulations \cite{peterson1995models, caccavale2014model, bayen2006adjoint, menon2006computer, yang2017fundamental, wei2013total, fleurquin2013systemic}, with different levels of granularity and hence of realism - from agent-based simulations, to data-informed models. These are generally based on virtual representations of the main ingredients and rules underpinning ATM, including sources of uncertainty and external factors (e.g. weather), yet avoiding unnecessary details; and aim at generating a trustworthy representation of how the system would have evolved under some given conditions. This allows to execute what-if analyses; e.g. a specific rule can be changed, and the consequences evaluated. Models and simulations have nevertheless to walk a delicate balance. On the one hand, too coarse-grained models often lack the capacity to fully represent the interactions within the system - as is the case of traditional queuing and Poisson‐based models~\cite{barabasi2005origin}. On the other hand, micro-scale simulations have to rely on data that are not always available; when these are substituted by estimations, the quality of the final results will strongly depend on the quality of these estimations.

A different and more recent approach is the one based on data-driven analyses: broadly speaking, instead of synthetically reproducing the dynamics of the system, they are based on analysing the data by it generated and on extracting information about the processes leading to a specific outcome. To illustrate this difference, one may create a model of an airport, encode rules about how flights avoid adverse weather events, and then evaluate their consequences \cite{janic2009modeling, kicinger2016airport, jen2022discrete, jones2025risk}; alternatively, one may consider a large number of days with different weather conditions, and extract relationships between these and observed delays \cite{schultz2018weather, zhou2020measuring, lui2022weather} \footnote{The frontier between both approaches is more fuzzy than what this simple example wants to illustrate; frequently, simulations are complemented by data-driven results, and the other way around.}. Data-driven analyses present the advantage of being hypothesis-free, in that they do not require a preconceived set of rules about how the system is going to react on a given situation - they only observe how the system actually reacted. Conversely, they do not easily allow to explore parts of the parameters' space not yet visited - i.e., if snow was never observed in a given airport, we cannot know how the system would react to it.

Within the vast family of data-driven analyses, a special place may be reserved for those based on statistical physics' concepts. Statistical physics is the branch of physics integrating elements of statistics and probability to study the macroscopic behaviour of systems with a very large number of microscopic constituents \cite{huang2009introduction, reichl2016modern}. As a prototypical example, imagine a gas. Studying it would prima facie require knowing the properties (position, velocity, etc.) of all particles composing it; this is clearly unfeasible. As an alternative, statistical physics allows describing it in terms of macro-scale properties, like temperature and pressure, that are derived by supposing the micro-scale ones follow some given statistical laws - hence, for instance, the temperature represents the average speed of all molecules. Within this context, physicists have developed many tools and techniques to infer the properties of individual elements within a system when only the large-scale, emergent dynamics are directly observable. Not surprisingly, these techniques have been used to study many real-world systems, from biological \cite{drossel2001biological, de2011contribution} to social \cite{castellano2009statistical} and technological \cite{chowdhury2000statistical, barabasi2001physics, pagani2013power, barthelemy2019statistical} ones. While more limited, ATM has also benefited from them \cite{cook2015applying}.

This review aims at presenting such techniques, specifically focusing on potential applications to the analysis of time series describing the dynamics of air transport and ATM; at discussing the information they can yield, as well as the limitations and challenges they pose; and at having an overview on what has already been done in the literature. Before delving deeper into the subject, let us review the logical progression in the analysis of complex systems from a statistical physics' perspective, as this will be the basis for the structure of this work.

\subsection{Statistical physics, from probability distributions to nonlinear dynamics}

The bedrock of any statistical physics approach to time series analysis is the identification of an empirical probability distribution, i.e. a sort of ``zero-th law", which grounds every subsequent measure of disorder via entropy‐based metrics. From this foundation, structural entropies quantify how unpredictably the system explores its range of states~\cite{tang2015complexity}. On the other hand, without the need of explicitly representing all possible micro-states, dynamical entropies offer a powerful approach to assess the temporal evolution of disorder. These entropies are not limited to static distributions but instead incorporate the sequential order of events, allowing the characterisation of how the system evolves over time~\cite{tang2015complexity}. This shift from static to temporal considerations is crucial for analysing real-world time series, where the order and duration of events carry essential dynamical information that static histograms might overlook.

In keeping with this time-oriented framework, tools from nonlinear physics emerge as a valuable option to unveil if that apparent randomness masks underlying determinism. Lyapunov exponents help detecting whether a system is sensitive to initial conditions, which is one of the fundamental ingredients towards chaos~\cite{kantz2003nonlinear} \footnote{Chaos describes the behaviour of deterministic systems that exhibit high sensitivity to initial conditions, leading to complex and seemingly random dynamics despite being governed by well-defined rules \cite{tsonis2012chaos}. For a visual example, one may consider a pool table: striking a ball with slightly different angles can lead to different bounces, and eventually to completely different final outcomes. This also has important implications for predicting the future dynamics of the system: in the presence of chaos, the temporal range for which forecasts are reliable is proportional to the accuracy with which the system is observed - as made famous by Edward Norton Lorenz in the case of weather \cite{lorenz2017deterministic}.}. Correlation dimension, on the other hand, estimates how many independent variables are truly shaping its behaviour~\cite{grassberger1983measuring}. Furthermore, multifractal analysis~\cite{kantelhardt2002multifractal} reveals hidden scale‐invariance and long‐range correlations, showing how bursts of activity cluster from the finest to the coarsest temporal scales. Last but not least, temporal irreversibility measures the asymmetry between forward and backward statistics, exposing the system's directional memory~\cite{weiss1975time,zanin2018assessing}. This concept is deeply connected to the notion of the arrow of time, that characterise how the statistical behaviour of large ensembles of individual entities of the system leads to irreversible macroscopic observations~\cite{wallace2013arrow}. 

The attentive reader would have noted a progression from low- to high-dimensional concepts: from considering the elements of the time series as independent; moving to grouping them into trajectories by including their appearance in sequences; to finally comparing trajectories in a non-linear way. This conceptual evolution is preserved in this review: we start from the basic (or even trivial) analysis of probability distributions (Sec. \ref{sec2}), for then moving to entropies (Sec. \ref{sec3}), and finishing in non-linear analyses (Secs. \ref{sec4} and \ref{NLDT}).

\subsection{What to expect from this review}

As we hope to make clear throughout this work, uncovering the dynamical properties embedded within air traffic time series, such as multifractality or chaoticity, is not merely an academic exercise. On the contrary, it is a crucial step towards understanding the true nature of the highly complex and interconnected air transport system. On the one hand, these properties reflect nonlinearity, non-Gaussian features, memory effects, and asymmetries, which are fundamentally incompatible with traditional Gaussian random models~\cite{9853327,wang2022flight,wang2022distribution}. From a practical perspective, this implies that models using estimations or synthetic versions of these time series must incorporate such properties.
On the other hand, these dynamical properties also yield insights about the mechanisms generating the time series, insights that are usually hidden and difficult to numerically prove.

If these are the advantages, the practitioner has also to be aware of the limitations and idiosyncrasies of such concepts. Consequently, each section below will be organised in three parts: an initial theoretical introduction, in which the fundamentals will be explained; a set of applications, i.e. examples of publications that have used such concepts in air transport; and concluding remarks about challenges and open directions. To facilitate the comprehension by readers without a statistical physics' background, its basic concepts will be explained through a series of footnotes. Finally, Sec. \ref{NLDT} will conclude by discussing some additional techniques that, while hitherto not used in air transport, may be the basis of future works.

As it can easily be imagined, some basic concepts (e.g. entropy) have widely been used already; presenting a review of all publications mentioning them would be unfeasible. In this work, we instead focus on selecting papers that are either historically important or relevant in terms of their findings. The early studies first brought statistical physics methods into air traffic data analysis, providing a foundational context. At the same time, we also consider studies that reported notable results, either through methodological innovation, by demonstrating how these theoretical tools could be employed as efficiency indicators, or by revealing potential applicability for evaluating the system's performance. This dual criterion ensures a well-rounded view that covers both the origins and the progress of this approach; still, if a given work is here not cited, it must not be taken as a undervaluation from our side. It is worth noting that this review is intended primarily for researchers in the field of aeronautics, rather than physicists. Our goal is to make the concepts and tools from statistical physics easier to understand and use when it comes to the evaluation and potential optimisation of operations. 

As a final note, while this review is focused on air transport and ATM, many of the concepts here discussed have applicability in other transportation modes, including rail, maritime, or even pedestrian mobility. They may further be relevant to transportation modes adjacent to air transport, such as urban air mobility \cite{straubinger2020overview, cohen2021urban} and unmanned aviation \cite{floreano2015science}. We hope this review will be a source of inspiration and insights also for these research communities.

\section{The very first step: probability distributions}\label{sec2}

\subsection{Empirical probability distributions}

In the statistical physics approach to time‐series analysis, the accurate estimation of the empirical probability distribution of an observable is the vital first step~\cite{jaynes1957information,beck2003superstatistics}. Just as the canonical distribution \footnote{The canonical or Gibbs' distribution describes the probabilities of finding a statistical system at equilibrium in any one of its stationary microscopic states. In other words, given a system with a specific macrostate (e.g. with a given temperature), the canonical distribution defines the probability of finding its constituents in a given configuration (e.g. the distribution of kinetic energy of all molecules). } in thermodynamics encodes all macroscopic properties of a system, filling the gap between microscopic disorder and the macroscopic outcome \footnote{Generally speaking, microscale refers to the individual elements composing a system, while the macroscale represents what is observed when they are taken together. To illustrate, the microscale of a gas would include all its individual particles, as well as all associated properties (as position, speed, etc.). Individual constituents are nevertheless disregarded when we consider a gas as a whole; the focus instead moves to global properties like temperature, pressure, and so forth. }, here it determines averages and fluctuation magnitudes through the moments, and any entropy-based disorder metric that can be evaluated. Think of a box of gas, where each molecule can have a range of energies, yet their exact energy at a given moment is unknown. What we do know is the probability distribution over those energies. From that distribution, we can compute the average energy, its variance (related to the heat capacity), and the state of disorder of the gas, i.e., its entropy. Finding the proper probability distribution that maximises that entropy subject to some constraints is exactly how statistical mechanics predicts equilibrium behaviour~\cite{jaynes1957information}. 

Let us analyse the activity of one airport through two basic statistical physics concepts: the microstate and the macrostate \footnote{A microstate refers to one of the many specific configurations of the system's individual components (e.g., positions and velocities of particles) that can give rise to the observable same outcome. A macrostate describes the overall, observable state of a system (e.g., temperature, pressure).}.
The former is the complete snapshot of every individual flight's status over a temporal interval: the exact landing and take-off times, taxi-in and taxi-out times, and the delay of the aircraft. On the other hand, the latter would be any aggregated time series of those details, as e.g. the average delay in one-hour intervals \footnote{It will be easy for the attentive reader to map the previous concepts to this example. To illustrate, a microstate corresponds to a specific set of operations that could have resulted in the observed macrostate, i.e. in a given delay evolution. Given a macrostate, the corresponding canonical distribution would tell us the probability that a given microstate was the cause of the observed macrostate. It could further be used to answer questions like: if we observe $15$ minutes of average landing delay, what is the probability that such average delay was caused by a single delayed flight?}. 
Let us start by the latter one, i.e. the average delay in a given time window. That first moment is nothing more than the center of mass of the delay landscape: it tells that, on average, each hour carries a given number of minutes of delays. If one is planning staff or gate assignments, that number is the baseline expectation. The second moment of those hourly delays, i.e. its variance, can be understood as an effective temperature for the operations. High variance warns that schedules and forecasts are unreliable, and may wildly mispredict the actual performance. One can progress to higher moments; to illustrate, a strongly positive skewness of the delay distribution indicates that large positive deviations dominate, which means that the airport rarely snaps back from congestion in a single hour.
In operational terms, these moments transform raw distributions into a physical portrait of the airport's dynamics: the first moment sets the operating point, the second diagnoses its stability, and the third uncovers directional bias toward congestion or recovery.

Moments, as previously described, are only the first approximation to ways of collapsing the ``raw state counting'' into a single measure. Among the alternatives, of special relevance are Shannon's~\cite{shannon1948mathematical} and other generalised entropies ~\cite{renyi1961measures, tsallis1988possible}. These measure the uncertainty or disorder in the system; or, in more technical terms, how many effectively accessible states the system can explore, just as in thermodynamics entropies represent how many microstates correspond to a given macrostate. This will be addressed in depth in the next section.

\begin{figure}[!tb]
\includegraphics[width=0.9\textwidth]{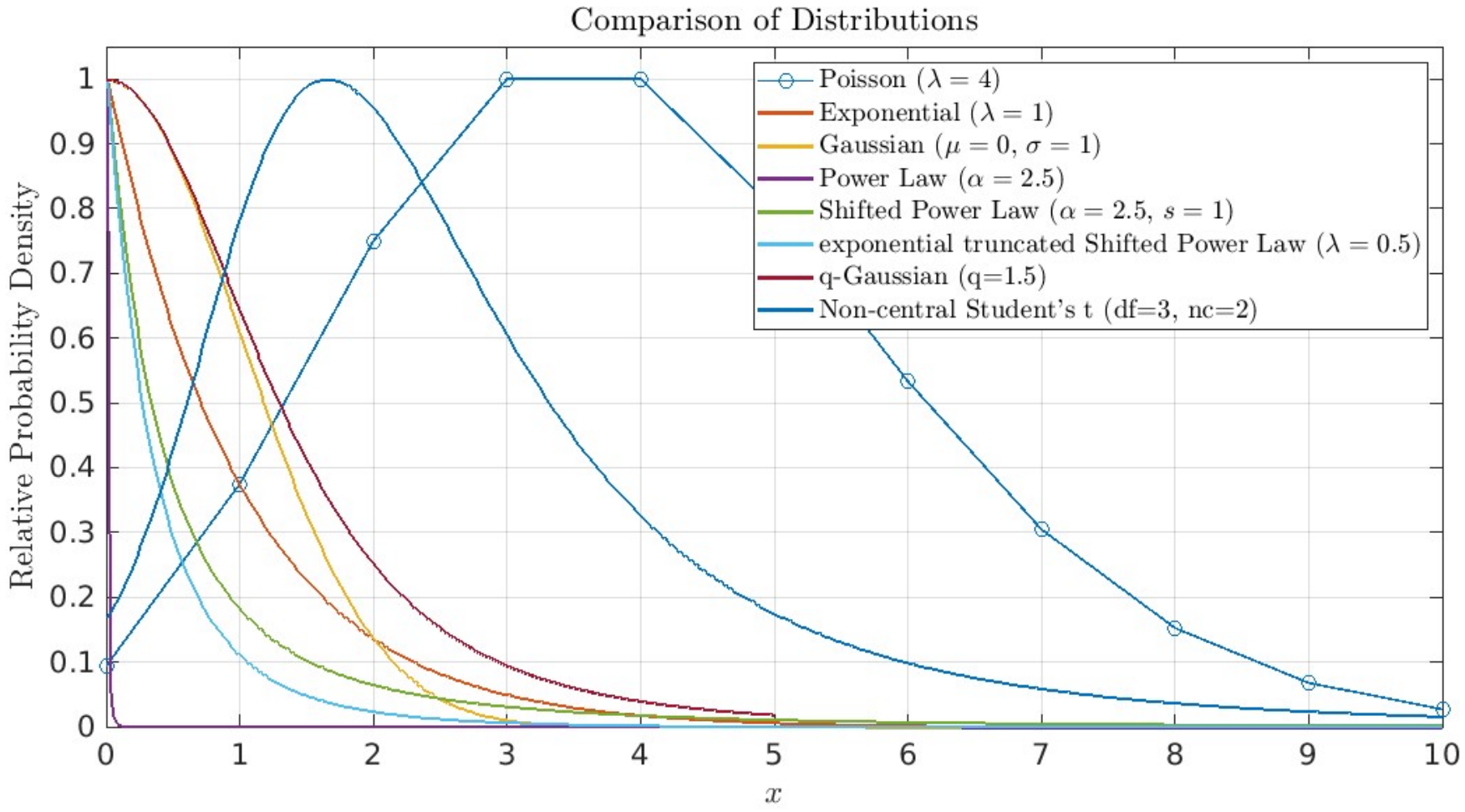}
\caption{ Graphical representation of the main theoretical distributions encountered in Sec. \ref{sec2}. Note how each family has different properties, e.g. asymmetry or decay rate, which are associated to properties of the dynamical process generating the data.
\label{fig:PDFs}}
\end{figure}

At this stage, it is important to clarify the difference between two concepts that are at time misunderstood. Upon measurement of the data, an empirical distribution describes the probability distribution inferred directly from the measurements~\cite{dekking2005modern}, and it is built by determining how frequently each individual outcome or collection of outcomes happens within a finite time series. To illustrate, if we are looking at flight delays, the empirical distribution is given by the histogram revealing how frequently each delay magnitude is observed. This yields a data-driven characterisation of the behaviour of the system without imposing any external assumptions. Alternatively, a theoretical distribution is one that is mathematically formulated to characterise the underlying process that has produced the observed data. Common examples include the Gaussian, exponential, and power-law distributions, each one corresponding to prior assumptions regarding the mechanisms generating the data. Theoretical distributions enable the derivation of analytical expressions for probabilities, moments, and other statistical values, so that predictions can be made beyond the observed data. 

The main difference between these two approaches is their derivation and use. Empirical distributions are essentially descriptive and outline empirically observed characteristics of the measurements in a non intrusive manner. They are a faithful representation of the measured frequencies, yet they are susceptible to sampling variability. Theoretical distributions present an idealised form that is designed to generalise and explain the observed behaviour within the data; however, their utility is dependent on the validity of the corresponding assumptions. A common practice in statistical analysis is to fit a theoretical distribution to an empirical one and quantify the ``degree of agreement'' by testing the goodness of fit~\cite{casella2024statistical}. This quantification indicates whether the theoretical model is a satisfactory approximation of the observed behaviour, and identifies systematical differences that may reveal additional structures in the measured process. 

One should pay attention to the fact that the universe of possible theoretical distributions is vast~\cite{forbes2011statistical}: there exist countless families of probability models, each with different assumptions and functional forms, ranging from classical distributions such as the Gaussian and exponential, to more specialised distributions such as the Student's t or stable distributions - a graphical depiction of the main families that will be encountered below is reported in Fig. \ref{fig:PDFs}. Such theoretical choices again emphasise the role of empirical distributions as a model-free representation of what has been observed, against which any theoretical suggestion is to be judged and verified: the more precise an empirical distribution is defined, the more solid any prediction or model will be~\cite{9853327, mitsokapas2021statistical}.

\subsection{Applications}

In what follows we present a non-comprehensive overview of the literature, focusing on the characterisation of empirical probability distributions in air transport, and especially highlighting the limitations of Gaussian-based assumptions in this context. For the sake of synthesis, these contributions are also listed in Tab.~\ref{tab:PDF}

Back in 2002, Muller {\it et al.}~\cite{mueller2002analysis} were among the first to systematically analyse delay statistics, examining 21 days of data from ten major US airports; they showed that departure delays conform to a Poisson law, while en‐route and arrival delays follow a Gaussian distribution. Later, considering the delays distributions from Denver airport, Tu {\it et al.}~\cite{tu2008estimating} showed that a mixture of four normal distributions best describes the departure delay distribution, with its pronounced right skew and heavy tails reflecting diverse operational and stochastic drivers.

In their 2004 study, Willemain {\it et al.}~\cite{willemain2004statistical} analysed the intervals between the estimated time of arrivals computed when the arriving aircraft is $100$ nautical miles from the destination, as the ``raw material" with which the final en-route and approach controllers must work to shape a more orderly arrival flow. They found a nearly exponential distribution of this arrival dataset in the nine major US airports. Consequently, as aircraft approach their destination, the interplay between airline schedules and en-route air traffic control creates a nearly random arrival pattern. This randomness imposes a significant workload on the controllers responsible for managing arrivals within the last 100 miles of flight. This fact was later also proven using metrics like fractal analysis and irreversibility~\cite{olivares2023measuring}. This result establishes that the exponential distribution is a meaningful benchmark against which the disorder of the stream can be measured. Later, it was empirically observed that the time intervals between successive aircraft landings at three major European airports follow a power-law scaling that applies only for times larger than two and smaller than 100 minutes~\cite{olivares2022corrupted}. On top of that, linear correlations were found in the daily arrival stream, which naturally entails the idea that the aircraft stream is successively rearranged to meet the airport control needs, producing interactions on the Poissonian process~\cite{caccavale2014model}. Power law decay in the time intervals between successive landings was also observed by Ito and Nishinari~\cite{ito2015universal} at hub US airports.

Next, Wang {\it et al.}~\cite{wang2020universal} examined 20 years of US data across 14 major airlines to reveal two universal clusters of departure delay propagation. By analysing the complementary cumulative distributions of propagated delay, they find one group whose delays follow a shifted power law, and another fitting an exponentially truncated shifted power law. These distributions quantify how likely is for a delay of size $L$ to propagate. The pure power law implies scale-free propagation with no characteristic delay size, while the exponential cutoff reflects operational limits that suppress extremely large cascades. By focusing on the distributions' shape, the authors demonstrate that the functional forms of delay propagation are remarkably consistent across years, suggesting an underlying universal process. A shifted power decay was also observed in departure delay distributions for Delta Airline~\cite{cao2019method}.

Across several UK airports, Mitsokapas {\it et al.}~\cite{mitsokapas2021statistical} showed that early arrivals (negative delays) exhibit approximate exponential decay, whereas late arrivals (positive delays) display a long‐tailed, $q$-exponential power‐law decay. Notably, they observe that all studied airports and even individual airlines at each airport exhibit a qualitatively similar distribution. The local $q$-exponential distribution with heavy tails seems to result from the superposition of many exponential distributions, i.e. to a superstatistics \footnote{Superstatistics refers to the process of superposing multiple differing statistical models to achieve the desired non-linearity, or the observed behaviour of the system \cite{beck2003superstatistics}.}. One can establish a benchmark against which to systematically quantify the delay performance by setting the optimal exponents that characterise the mentioned decays. Heavy tails were also observed in empirical distributions of per-flight departure and arrival delays~\cite{fleurquin2013systemic}. According to a more recent analysis by Z. Szab\'o~\cite{szabonon}, the empirical delay distributions in Europe and the US are well described by a non-central Student's t distribution. Finally, multiple probability distributions for modelling flight delays at Guangzhou Baiyun International Airport were compared, including Beta, Erlang, and Normal distributions~\cite{wang2022distribution}; results showed that the Gaussian one is the best at capturing delay stochasticity. 

Recognising that European and US flight departure delay distributions deviate from a Gaussian behaviour, a recent study~\cite{olivares2025quantifying} has used a quantitative measure~\cite{zunino2022permutation}, based on ordinal patterns~\cite{bandt2002permutation}, for quantifying their skewness. The authors found that high-traffic hubs consistently exhibit larger departures from normality, with US airports showing predominantly negative skewness (favouring early arrivals) and European airports displaying more heterogeneous profiles. Seasonal and airport-specific analyses further illustrated how congestion and structural changes (e.g., airline shifts or capacity reductions) influence the shape of delay distributions.

\subsection{Lessons learned and future directions}

It may be tempting to assume that a detailed characterisation of a probability distribution is a nice but irrelevant exercise. While this may be the case in other fields of science, it does not hold true in the context of statistical physics. In place of begin with assumptions (e.g., Gaussian noise, Poisson arrivals), this framework insists on letting the data speak first; the empirical distribution thus becomes the model in its rawest form.

The importance of a correct characterisation of the empirical distribution is two-fold. On the one hand, it allows to correctly encode the behaviour of the system; to derive moments (e.g. mean, variance, skewness) and any other distribution-based metric; and further to describe their evolution through time. It thus supports any following modelling and analysis task. For instance, in Sec. \ref{NLDT} we will introduce the concept of the Hurst exponent, whose estimation can substantially be biased by the presence of heavy tails.
On the other hand, and as seen in previous cases \cite{willemain2004statistical, caccavale2014model, wang2020universal}, the shape of the distribution itself can be used to describe the mechanisms generating those data: Gaussian, scale-free and cutoffs may respectively point to the appearance of random delays, to propagations thereof, and to operational limits. Superstatistical frameworks \cite{beck2003superstatistics} may thus be required to fully describe the dynamics of the system.

The attentive reader will also have noted that delays have been described through many different distributions: from Gaussian \cite{mueller2002analysis, tu2008estimating, wang2022distribution}, to Poisson \cite{mueller2002analysis}, exponentials \cite{mitsokapas2021statistical}, Student's t \cite{szabonon}, and power laws \cite{wang2020universal, mitsokapas2021statistical}. This may stem from two causes. Firstly, delays are not always defined in the same way. To illustrate, depending on the data source and how they are calculated, they may refer to the operational (i.e. scheduled landing time) or commercial time of arrival; they may use the initial plan of the airline, or the one updated according to Air Traffic Flow Management constraints (as is the case of the EUROCONTROL's R\&D Archive); and may even have different levels of granularity \footnote{For a comparison of data sources across regions, the interested reader may refer to Ref. \cite{cook2017atm}.}. Secondly, fitting a given data set to a distribution, and especially choosing between different models, is not a trivial task, and is even more challenging when non-linearities are present \footnote{For a long-standing discussion of whether real-world systems are really scale-free or not, the interested reader can refer to Refs. \cite{lima2009powerful, broido2019scale, voitalov2019scale, smith2021scarcity}.}. In short, it may be concluded that even this first, and prima facie trivial, step has not completely been solved.


\begin{center}
\begin{table}[h!]
\begin{tabular}{|c| p{4cm} |p{0.5cm} |p{2.2cm}|p{1.5cm} |p{1.5cm}|} 
 \hline
Year & Title & Ref. & Geographic scope & Temporal scope &Type of data \\ [0.5ex] 
 \hline\hline
  2002 & Analysis of aircraft arrival and departure delay characteristics & \cite{mueller2002analysis} & US & Oct. 14th - Nov 3th 2001 & Dep. delays \\ 
 \hline
 2004 & Statistical Analysis of Intervals between Projected Airport Arrivals & \cite{willemain2004statistical} & US & Dec. 2003 & Time intervals between ETA$^{*}$  at 100M \\ 
 \hline
  2008 & Estimating flight departure delay distributions—a statistical approach with long-term trend and short-term pattern & \cite{tu2008estimating} & US (United Airlines) & 2000-2001 & Dep. delays \\ 
 \hline
   2013 & Systemic delay propagation in the US airport network & \cite{fleurquin2013systemic} & US & 2010 & Dep. and arr. delays \\ 
 \hline
    2015 & Universal bursty behavior in the air transportation system & \cite{ito2015universal} & US & 2014 & Time intervals between landing \\ 
 \hline
   2019 & A method of reducing flight delay by exploring internal mechanism of flight delays & \cite{cao2019method} & US (Delta Airlines) & July to Dec. 2017 & Dep. delays \\ 
 \hline
   2020 & Universal patterns in passenger flight departure delays & \cite{wang2020universal} & US & 1995-2015 & Dep. delays. \\ 
 \hline
  2021 & Statistical characterization of airplane delays & \cite{mitsokapas2021statistical} & UK & 2018-2020 & Arr. delays. \\ 
 \hline
   2022 & Corrupted bifractal features in finite uncorrelated power-law distributed data & \cite{olivares2022corrupted} & Europe & May 2018 to July 2021 & Time intervals between landing \\ 
 \hline
    2022 & Distribution prediction of strategic flight delays via machine learning methods & \cite{wang2022distribution} & China (Guangzhou Baiyun Airport) & March 26th 2017 to March 28th 2020 & Arr. and Dep. delays \\ 
     \hline
    2023 & Non-linear transitions in air transport delays: models and data & \cite{szabonon} & Europe and US & 2015-2022$^{**}$ & Arr. and Dep. delays \\ 
         \hline
    2025 & Quantifying deviations from Gaussianity with application to flight delay distributions & \cite{olivares2025quantifying} & Europe and US & 2015-2019 & Dep. delays \\ 
 \hline
\end{tabular}
\caption{List of all papers in chronological order analysing the probability distribution of different air traffic data. ($*$) ETA: Estimated Time of Arrival. $(**)$ For Europe, only Mondays between 08/12/2014 and 27/02/2023 were considered.}
\label{tab:PDF}
\end{table}
\end{center}

\section{Entropy-based metrics}\label{sec3}

Entropy has emerged as a fundamental concept for quantifying the disorder in time series and, in some cases, their complexity~\cite{pincus1991approximate,bandt2002permutation,wang2024complexity} \footnote{Note that, in statistical physics, disorder and complexity are two different concepts, even though at times they are mixed together. A deeper discussion will be provided in Sec. \ref{sec:future_topics}.}. In the context of air‐traffic operations, disorder refers to the degree of unpredictability or irregularity in key performance indicators. This is most intuitively seen in an airport's individual arrival delay sequence. Imagine two consecutive days at a busy hub: on day 1, the delay of each individual arrival remains constant throughout the day, then we have a complete knowledge of what is happening. That is a highly ``ordered" dynamics, where incoming delays are perfectly anticipated from the past trend. Now consider day 2, when internal (operational) constraints together with external factors combine to produce a jagged arrival delay sequence: one landing arrives with 10 minutes of delay, the next with 45 minutes, then back down to 5 minutes with the third one, then up to 60 minutes, with no clear patterns. This erratic behaviour embodies high disorder. 

Formally, disorder can be defined as the departure from a simple, low‐dimensional pattern (like the one having constant arrival delays) toward a complex, high‐dimensional one with many competing influences. More specifically, disorder may arise when external factors (e.g. bad weather, en-route ATC constraints) interact non-linearly with operational constraints (late boarding due to aircraft and crew availability), creating feedback loops that break the equilibrium between busy and quiet periods. The more these loops dominate, the more jagged the sequence of delays and, consequently, the more irregular the arrival and departure flows become. By quantifying this disorder through measures like standard deviation or entropy-based metrics of aggregated sequences, one can distinguish days or airports operating in an orderly, manageable regime, from those in a disordered or complex state. Recognising when disorder crosses a critical threshold enables proactive interventions (such as dynamic flow control, additional staffing, or weather‐related contingency plans) to restore smoother, more predictable operations.

\subsection{Structural versus dynamical entropy}

Entropy, in its classical formulation introduced by Claude Shannon~\cite{shannon1948mathematical}, provides a fundamental measure of the disorder associated with a probability distribution \footnote{A concept similar to Shannon's entropy was previously introduced in statistical physics, firstly by Ludwig Bolzmann in 1866, and later generalised by Josiah Willard Gibbs in 1878 \cite{jaynes1965gibbs}. In this thermodynamics' interpretation, entropy represents the degree to which the probability of a system, as observed in its macrostate, is spread out over different possible microstates.}. In its discrete version, Shannon's entropy is defined as:

\begin{equation}
    S = - \sum_i p_i \log p_i.
\end{equation}

\noindent where $p_i$ is the probability of the symbol $i$, and $i$ runs over all possible symbols.
Intuitively, this definition can be understood as a measure of how spread out (or uncertain) the distribution is, and quantifies the average information required to specify an outcome of a random variable. This concept has been instrumental not only in Information Theory~\cite{shannon1948mathematical} but also in the study of complex systems, where entropy serves as a bridge between microscopic variability and macroscopic order~\cite{jaynes1957information}.

Building upon this foundation, it is possible to categorise structural and dynamical forms of entropy, each capturing different aspects of the system behaviour. The former evaluates the disorder of a system through the distribution of ``energy" or power across different frequencies. Techniques such as spectral~\cite{powell1979spectral} and wavelet entropies~\cite{rosso2001wavelet} transform a time series into the frequency or time-frequency domain, for then assessing the uniformity or concentration of power. A highly concentrated spectrum (e.g., a single dominant frequency) indicates low entropy and high regularity, while a broader spectrum suggests greater disorder. Structural entropy is particularly useful in identifying periodic and global memory, yet it is less sensitive to local changes in temporal correlations.

In contrast, dynamical entropy captures local temporal structures by assessing how often patterns repeat and how predictable future states are. Measures such as the Approximate~\cite{pincus1991approximate} and Sample entropies~\cite{richman2000physiological} quantify the likelihood that initially similar embedding vectors remain similar over time, reflecting the degree of randomness or determinism in the signal. More specifically, both entropies begin by transforming a time series into a set of $m$-dimensional vectors, and, for each of them, count how many other vectors lie within a distance $r$ (also called the tolerance). Finally, both entropies compute the logarithmic likelihood that nearby $m$-length matches remain close when extended to length $m+1$. The main difference between both is that the Approximate entropy includes self‐matches, which can bias the estimation, particularly for short or noisy records; on the other hand, Sample entropy excludes self‐comparisons, yielding a more consistent estimation of the likelihood, and making it less sensitive to record length and more robust against noise~\cite{richman2000physiological}. Even though both metrics require a careful selection of the embedding dimension $m$ and tolerance $r$~\cite{delgado2019approximate}, they have successfully been employed on a wide range of data collected from diverse scientific areas~\cite{staniek2008symbolic, lake2002sample, yentes2013appropriate, chou2014complexity, xavier2019application, olbrys2022approximate}. For a more detailed methodological discussion of these two entropic measures, the interested reader can refer to Ref. \cite{tang2015complexity}.

Bridging the structural and dynamical features in time series, permutation entropy offers a concise yet powerful framework that integrates both structural variability and dynamical information~\cite{bandt2002permutation}. From a structural perspective, this entropy captures the diversity of ordinal patterns, i.e. the order required to rank values in short segments of the time series; such diversity reflects the range of possible configurations the system can adopt, akin to how Shannon's entropy evaluates the diversity in a histogram~\cite{martin2006generalized}. Simultaneously, ordinal patterns inherently encode temporal organisation: the ordering of sampled values embodies the local dynamics and the sequential dependencies between observations~\cite{zunino2012distinguishing}. This makes permutation entropy sensitive to the temporal structure of the time series, including its correlations~\cite{zunino2008permutation}, deterministic trends~\cite{rosso2007distinguishing}, forbidden patterns~\cite{amigo2007true} and motifs that would be invisible to purely static (structural) metrics~\cite{martin2006generalized,parlitz2012classifying}. This ability to capture, in tandem, the structural diversity and the temporal dynamics in a time series is precisely what has made permutation entropy so widely adopted across numerous fields, not least air transport.

\subsection{Applications}

Tab.~\ref{tab:entropic} presents an overview of the literature that has applied the concept of entropy for characterising the complexity of aggregated time series from air traffic.

The first attempt at quantifying the complexity of an airport dates back to 2012, led by Dong and Du, who applied approximate entropy to characterise air traffic flows in the terminal area~\cite{dong2012analysis}. They measured the entropy of the time differences between departure and arrival flights, for then comparing these values with those derived from chaotic sequences (specifically, the Logistic\footnote{The Logistic map \cite{sprott2003chaos} is a simple yet powerful mathematical model used to describe how populations grow over time under limited resources. It is a one-dimensional recursive equation that relates the population at one time step to the next. When the growth rate parameter is increased beyond a certain threshold, the system exhibits complex behaviour, including bifurcations and chaos. Despite its simplicity, the Logistic map has become a key example of how deterministic systems can produce unpredictable, seemingly random dynamics.} and H\'enon\footnote{The H\'enon map \cite{sprott2003chaos} is a two-dimensional discrete-time dynamical system introduced by Michel H\'enon in 1976 as a simplified model of chaotic behaviour in a dissipative system. It consists of a pair of recursive equations that generate a sequence of points in the plane. Despite its simple form, the map produces a fractal structure known as the H\'enon attractor, which illustrates how deterministic rules can lead to complex dynamics in the phase space.} maps) and a completely random sequence. The results indicated similarities between the experimental data and the chaotic maps, leading them to suggest that air traffic flows might exhibit chaotic behaviours. Later, Wang and co-workers applied the sample entropy and the multi-scale sample entropy to measure the complexity of air traffic flow~\cite{wang2019complexity}, focusing on $28$ days of operations in three sectors of the Sanya's airspace, China. Entropic values remained stable at larger time scales and showed low dependence on time series length, making them suitable for real-time dynamic monitoring. Traffic flows for longer time intervals (e.g., 60 minutes) lead to entropy increment, indicating greater randomness and reduced predictability. This suggests that short-term air traffic flow prediction is feasible, while long-term forecasting is more challenging.

Liu and co-workers~\cite{liu2020multiscaleA} proposed a method for quantifying the complexity of the air traffic flow of the ten busiest airports in China. They introduced an improved multivariate multiscale permutation entropy, which allows for the simultaneous analysis of multi-channel data to measure the complexity of airport traffic flow fluctuations. Studying how this entropy metric evolves throughout the day makes it possible to assess all the characteristic temporal scales of the traffic flow dynamics, with the most significant scale being the one of six hours. Particularly, for arrival, departure, and total traffic volume, they observed several drops in the entropy as the temporal scales changed; the number of these drops was then used to group airports with similar dynamics. In order to understand the rationale behind this analysis, it has to be highlighted that representing a time series through ordinal patterns for different lags $\tau$ is equivalent to ask how does the system look when it is sampled every $\tau$ steps. In a purely random process, all patterns are equally likely and the permutation entropy is maximal and independent on $\tau$. However, for a periodic (or quasi-periodic or time-delayed) system, when $\tau$ matches such period, ordinal patterns always represent the same phase of the cycle. Every pattern is therefore perfectly repeating, yielding a distribution far from uniform and a lower permutation entropy~\cite{zunino2010permutation}. This does not only hold for the period itself, but also for harmonics (i.e. multiples) and sub-multiples of it. Accordingly, these results unveil the periodic clockwork of an airport dynamics rather than any complex feature of the flow dynamics. 

The multivariate approach proposed in Ref. \cite{liu2020multiscaleA}, i.e. considering both arrival and departure data as two complementary time series, presents the advantage of reducing noisy entropy fluctuations; still, amplitude information is lost in the ordinal pattern representation~\cite{bandt2002permutation}. This latter limitation motivated the same team to propose an improved multivariate multiscale weighted permutation entropy~\cite{liu2022exploring}, obtaining a similar qualitative characterisation. Even though these studies provide a detailed estimation of a complexity measure for volume flows, there is a limited connection between the findings and the operational characteristics of the airports. 

A completely different approach was introduced in Ref.~\cite{olivares2023markov}, based on the representation of the hourly arrival volume through ordinal patterns and on the study of the resulting probability distributions - i.e. the whole distribution, as opposed to the corresponding entropy. These distributions present different features for airports with one and two runways dedicated to landings. Moreover, minimising the distance between these and the one obtained from a Markov-modulated noise allows the estimation of the correlation between consecutive hours in the arrival flow, which can be interpreted as a metric of efficiency. A comparison of the dynamics pre and post COVID-19 illustrated that the reduction of randomness seen post-pandemic is not solely attributable to decreased traffic; rather, the correlations diminished more significantly than anticipated, i.e. if aircraft interactions would have remained constant despite the traffic volume.

It is further worth considering the work of Martinez {\it et al.}~\cite{martinez2023complementarity}; following the recipe of building informational planes, they introduced the entropy–time asymmetry plane using ordinal patterns-based metrics. The analysis of arrival delay data at 12 major European airports revealed that most airports show high permutation entropy values. This result suggests that delays are essentially random rather than following predictable patterns, with the level of randomness depending on the size of the airport.

\subsection{Lessons learned and future directions}

The use of entropy as a proxy for complexity remains the subject of active debate. While entropy-based metrics effectively quantify disorder or unpredictability of a system, they do not necessarily measure structural complexity. For instance, purely random sequences (i.e. white noise) typically yield high entropy values, yet are often considered low in complexity due to the absence of significant structures. Conversely, deterministic systems, such as those exhibiting chaotic behaviour, may have medium/low entropy, while still generating complex temporal structures~\cite{feldman1998measures}. This paradox highlights the limitation of using entropy on its own: high entropy does not imply high complexity, nor medium/low entropy always stands for simplicity. Consequently, many researchers opt to combine entropy with other metrics, such as statistical complexity or non-linear tools, to achieve a more accurate characterisation of the underlying dynamics~\cite{rosso2007distinguishing}. 

In particular, permutation entropy has proven to be a powerful tool for analysing complex temporal dynamics~\cite{zunino2010permutation}. One of the most compelling advantages of permutation entropy is the ability of identify characteristic temporal scales in time series by analysing how the entropy varies with the sampling time, i.e. the lag $\tau$, as evidenced in the case of air traffic flow~\cite{liu2020multiscaleA,liu2022exploring}. In this context, Zunino {\it et al.}~\cite{zunino2011characterizing} have shown that in the case of a chaotic systems with delayed feedback, the permutation entropy evaluated at the feedback delay qualitatively reproduces results consistent with the classical Kolmogorov-Sinai entropy, thereby capturing the chaotic nature without requiring the estimation of the Lyapunov exponent, which is often challenging in experimental scenarios. In the light of these results, a promising path opens for applying this approach to air traffic flow analyses. Specifically, the presence of those local minima in the permutation entropy, already observed by Liu {\it et al.}~\cite{liu2020multiscaleA,liu2022exploring}, suggests intrinsic temporal scales within the airport dynamics. This approach could offer a new perspective for understanding congestion and delay propagation.

\begin{center}
\begin{table}[h!]
\begin{tabular}{|c| p{4cm} |p{0.5cm} |p{2.2cm}|p{1.5cm} |p{1.5cm}|} 
 \hline
Year & Title & Ref. & Geographic scope & Temporal scope &Type of data \\ [0.5ex] 
 \hline\hline
  2012 & Analysis of complexity measure of air traffic flow at terminal area based on approximate entropy & \cite{dong2012analysis} & Unspecified & Unspecified & Time difference between dep. and arr. (at runway) \\ 
 \hline
   2019 & Complexity analysis of air traffic flow based on sample entropy & \cite{wang2019complexity} & China & 1th-28th Oct. 2007 & Vol. \\ 
 \hline
 2020 & Multiscale complexity analysis on airport air traffic flow volume time series & \cite{liu2020multiscaleA} & China & summer 2017 & Arr., dep. and total Vol. \\ 
 \hline
  2022 & Exploring the impact of flow values on multiscale complexity quantification of airport flight flow fluctuations. & \cite{liu2022exploring} & China & summer 2017 & Arr., dep. and total Vol. \\ 
 \hline
     2023 & Markov-modulated model for landing flow dynamics: An ordinal analysis validation & \cite{olivares2023markov} & Europe & 2018-2019 /2020-2021 & Arr. Vol. \\ 
 \hline
   2023 & On the complementarity of ordinal patterns-based entropy and time asymmetry metric & \cite{martinez2023complementarity} & Europe & Sep. 2018 & Arr. delay \\ 
 \hline
\end{tabular}
\caption{List of all papers in chronological order using entropic-like metrics for analysing different air traffic datasets.}
\label{tab:entropic}
\end{table}
\end{center}

\section{Fractality}\label{sec4}

Exploring real-world time series by uncovering their long-range correlations and scale-invariant (fractal) structures has become a foundational technique for describing empirical phenomena, including physiological records~\cite{zanin2022gait, peng1993long}, urban traffic~\cite{peng2010long, wang2014multiscale, xu2015traffic, thakur2015evidence, krause2017importance, feng2018better}, air traffic~\cite{olivares2022corrupted, olivares2023measuring}, atmospheric turbulence~\cite{funes2016synthesis, olivares2021high}, ocean dynamics~\cite{ozger2011scaling}, or pollution~\cite{he2016multifractal}. 
Fractality and long-range correlations are two sides of the same coin and emerge from scale-invariance, meaning that there is no single characteristic time scale that governs the process under analysis. In a fractal (self‐similar) time series, when one ``zooms out'' by grouping data into larger and larger blocks, the fluctuations look statistically the same. This lack of a characteristic time scale shows up in long-range correlations, where values separated by arbitrarily long lags $\tau$ remain statistically linked.

In a memory-less (Markovian) process, such as flipping a fair coin, knowing the last flip tells you nothing about the next one. Conversely, rainy/dry weather exhibits memory: both rainy days and dry stretches tend to cluster together. This ``memory'' is what constitutes long-range correlations: distant points in time remain linked, such that what happens now will still echo several steps in the future~\cite{beran2017statistics}. From the perspective of air transport operations, if delays were purely random, the hourly aggregated delays at an airport would be independent, like flipping a fair coin at every arrival. Nevertheless, given the finite capacity of airports, any external disturbance (such as bad weather, crew shortages, or airspace congestion) may postpone a group of landings, creating a ripple effect and delaying subsequent flights. What one observes in reality is the latter case, i.e. a persistent clustering: once the system slips into a backlog, those delays tend to persist and busy hours ``echo'' into the next ones rather than resetting to average. A similar argument can be made for the time intervals between successive landings at a busy airport, which are characterised by both series of very short gaps (i.e. clustered arrivals) and long pauses (i.e. when the flow eases). By plotting those inter-landing times at different temporal resolutions, i.e. by aggregating those intervals into larger blocks to get an increasingly coarser time series, fluctuating patterns of clusters and low-traffic phases will emerge at all temporal scales, illustrating the fractal nature of the dynamics.

Mathematically, this shows up in the autocorrelation function decaying slowly, as a power law, rather than exponentially~\cite{beran2017statistics}. The correlation between data points separated by a time gap $\tau$ decreases like $\tau^{-\gamma}$ for some exponent $\gamma$ between 0 and 1. To capture and compare the strength of these correlations, a widely used metric is the Hurst exponent, usually denoted by $H$~\cite{hurst1951long}. For many fractal processes, the exponent in the correlation function $\gamma$ and the Hurst's one are directly related, e.g. $\gamma = 2 - 2H$ in fractional Gaussian noises~\cite{kantelhardt2001detecting}. In the more general case, $H$ can be estimated by measuring how the characteristic fluctuation size $F(s)$ in windows of $s$ data points grows with $s$~\cite{peng1994mosaic}. In a fractal long-range correlated process, one finds, within a certain range of $s$: 
\begin{equation}
    F(s) \propto s^{H}.
\end{equation}
Positive memory or persistent clustering corresponds to $H>1/2$, negative memory or anti-persistence (where high values tend to be followed by lows) relates to $H<1/2$, while $H=1/2$ stands for a memory-less time series. In more intuitive terms, $H$ is thus a single number between $0$ and $1$ that summarises how ``sticky'' the past is.

Typically, only one scaling exponent $H$ is required to characterise the global linear correlations present in the data - a situation that is known as monofractality~\cite{kantelhardt2001detecting}. Nevertheless, in certain cases, the underlying dynamics is the result of an interplay between several elements, each with its own scaling due to non-linearities. In such instances, the scaling becomes a local property, and multiple scaling exponents $h$ are needed to properly characterise what is called multifractality~\cite{kantelhardt2002multifractal}. Even though heavy-tailed distributed data~\cite{kwapien2023genuine}, isolated singularities~\cite{oswiȩcimka2020wavelet} and finite-size effects~\cite{grech2013multifractal} can also lead to observe multifractal properties, these are considered as spurious, and genuine multifractality is only accepted when stemming from non-linear long-range temporal correlations~\cite{kwapien2023genuine}. In the absence of these correlations, only bifractality may occur~\cite{nakao2000multi}.

\subsection{Multifractal Detrended Fluctuation Analysis}

Researchers have developed several approaches to estimate the Hurst exponent, each with its own strengths and limitations. The earliest and simplest is Rescaled–Range $(R/S)$ analysis~\cite{hurst1951long}, in which one divides the time series into segments of increasing size $s$, computes the range of cumulative deviations $R$ from the mean in each window, and normalises them by the local standard deviation $S$. For fractal long-range correlated sequences, it can be shown that $\langle R/S \rangle \propto s^{H}$ (with $\langle \cdot \rangle$ denoting the average). While historically important, $R/S$ analysis can be biased by underlying trends or shifts in the data, leading to over– or under-estimations of the memory. To address these shortcomings, Detrended Fluctuation Analysis (DFA) was later introduced~\cite{peng1994mosaic}. It begins by integrating the series, for then, within each segment of size $s$, fitting and subtracting a local trend (often linear, quadratic, or cubic). The root–mean–square fluctuation $F(s)$ of the detrended data is then calculated across all window sizes; finally, a log-log plot of $F(s)$ as a function of $s$ should result in a straight line whose slope is $H$. 

Among the alternatives to estimate the Hurst exponent it is worth mentioning a family of frequency-domain ones~\cite{geweke1983estimation}, which extract the power spectral density $S(f)$ via Fourier transform and fit a low-frequency power law $S(f) \propto f^{\beta}$ to infer $H=(1+\beta)/2$. Specifically, wavelet-based estimators~\cite{abry1998wavelet} decompose the series into localised time-frequency bands whose coefficient variances scale as $2^{2jH}$, offering built-in detrending and superior localisation, at the cost of requiring the selection of an appropriate mother wavelet and scale range~\cite{abry1998wavelet}. In spite of these advantages, DFA has become the \emph{de facto} standard approach in many fields, as it strikes a great balance between computational simplicity and robustness to non-stationarities for quantifying long‐range dependencies for real‐world data~\cite{kantelhardt2001detecting}. 

Multifractal Detrended Fluctuation Analysis (MFDFA)~\cite{kantelhardt2002multifractal} extends the standard DFA framework by using a whole family of fluctuation functions $F(s)$ of different orders $q$; this allows to weight the contributions of both small and large deviations in the data, thereby uncovering a spectrum of scaling exponents $h(q)$ rather than a single exponent. For a monofractal sequence, $h(q)$ is independent of $q$ and equal to $H$, implying that the exponent $H$ is sufficient to characterise the dynamics. Conversely, in a multifractal scenario, $h(q)$ decreases with $q$, meaning that a full set of $\lbrace h_q\rbrace$ is required to describe the multiple scalings~\cite{kantelhardt2002multifractal}. In real-world measurements, the scaling properties are often dependent on the temporal scale used to analyse the multifractality; in other words, there may exist a crossover that separates two or more regimens, each with different scaling exponents. In such cases, to prevent subjective selection of crossover times, Giera{\l}towski {\it et al.} \cite{gieraltowski2012multiscale} introduced a multiscale generalisation of the MFDFA approach by defining a $h_q$-surface over different time scales. For further details about this methodology and its implementation, the interested reader may refer to Refs.~\cite{gulich2014criterion, thompson2016multifractal, ihlen2012introduction}.

\subsection{Applications}

To the best of our knowledge, Tab. \ref{tab:MF} lists all papers analysing fractal and multi-fractal properties of different air traffic data sets. In what follows we review the application of these concepts, organised by the type of data.

{\bf{Delay-related sequences}} ---  Long-range correlations were observed in six types of delay-related time series~\cite{lan2020characteristic}. More specifically, for one-hour intervals, Lan and Shangheng computed the aggregated total count of delays, their occurrence rate, and the average delay for both departures and arrivals. Note that a threshold was applied to delays, such that only deviations from the planned arrival and departure times larger than $15$ minutes were considered. For all sequences analysed, the Hurst exponent lied in the range $(0.5, 1)$, meaning that the number, the rate, and the average of delays for both arrivals and departures are positively correlated.

{\bf{Landing time intervals sequences}} --- To study the interactions between aircraft during landing, the authors of Ref.~\cite{olivares2023measuring} analysed the intervals between estimated landing times at the 12 and 10 major airports in Europe and China, respectively. They found long-range correlations, indicated by a Hurst exponent $H>1/2$, at both European and Chinese airports, even though the data were collected using different methodologies. Seasonal invariance of $H$ suggests that these temporal correlations determine systemic properties of airports---such as how aircraft are sequenced for landing---rather than simply reflecting the volume of traffic or specific weather conditions. Additionally, by contrasting the temporal correlations measured at 10.000 feet of altitude with those at landing, they showed that aircraft interactions mostly appear at the final approach phase. These findings align with what reported in other studies, as e.g. Ref.~\cite{willemain2004statistical}. Furthermore, it was demonstrated that the COVID-19 worldwide pandemic period displayed significantly smaller long-range correlations and, consequently, landing interactions. 

Multifractal properties were also found in landing time intervals at three major European airports (Frankfurt, Heathrow, and Tegel)~\cite{olivares2022corrupted}. The authors found a bifractal nature, indicating that the multiscaling is primarily characterised by longer time intervals rather than shorter ones, which are close to the minimum separation time. Daily estimation of the Hurst exponent showed that long-range linear correlations are present even at shorter temporal scales~\cite{olivares2022corrupted}.  

{\bf{Traffic flow volume}} --- To explore the temporal structure and memory characteristics of air traffic flow, Wang {\it et al.}~\cite{wang2018nonlinear} examined its scaling properties. At all examined intervals (10, 15, and 30 minutes) the Hurst exponents were consistently larger than $0.5$, indicating the presence of long-term positive correlation. This implies that increases or decreases in the traffic flow are likely to be followed by similar trends. Later, Zhang {\it et al.}~\cite{zhang2019multifractal} analysed multifractal air traffic flow volume properties, including arrivals, departures, and total volume in $5$-minute intervals. They found a highly persistent multifractal dynamics with $H \sim 1$ at small scales, along with a crossover scale of approximately $26$ hours for both arrival and total flow volumes, and $19$ hours for departure ones. This indicates that a set of scaling exponents characterises the fluctuation of the daily traffic flow; large volumes scale differently than small ones, and their strong correlation is the source of the multifractality. In contrast, when examining the dynamics over temporal windows larger than a day, the volume fluctuations resemble a monofractal anticorrelated noise, characterised by a simpler dynamics with only one scaling exponent $H = 0.18$. Additionally, it was found that weather conditions affect the multifractal properties of the flow volume, yet the results are heterogeneous; during thunderstorm season, multifractality decreases for both arrival and total flow volumes, while it increases for departure.

More recently, Liu {\it et al.} \cite{liu2020multiscaleB} took one step further by characterising the multiscale multifractal \cite{gieraltowski2012multiscale} dynamical properties by continuously changing the temporal scales on the departure flow volume. Their empirical findings demonstrated that a duration of only $101$ days is sufficient to explore the multifractal properties of the data. This result is vital for supporting studies using sliding windows to analyse the temporal evolution/transitions between seasons. Furthermore, they showed that for scales larger than 8 hours, multifractality is insensitive to fluctuation of large volumes and is dominated by fluctuations of small ones. This contrasts with the behaviour observed for scales smaller than 8 hours. Despite the detailed nature of these results, a lack of interpretability in terms of real-world operations still remains.

\subsection{Lessons learned and future directions}

Fractal-based techniques are commonly used to asses long-range correlations in times series, especially in terms of their Hurst exponent. Nevertheless, its interpretation requires caution, particularly when the dynamics present clear oscillatory patterns. In such cases, the oscillation can artificially increase the value of $H$, leading to a overestimation of the long-range correlations~\cite{katsev2003hurst, ludescher2011spurious}. This is particularly relevant on real-world measurements, like air traffic data sets, where daily trends are an inherent part of the dynamics. 

Periodic or quasi-periodic components can distort the scaling behaviour detected by methods like MFDFA, leading to a misinterpretation of the multifractal properties of the system. Particularly, oscillatory trends can introduce artificial crossovers in the fluctuation function, mimicking multifractality and falsely suggesting long-range correlations~\cite{katsev2003hurst,ludescher2011spurious,olivares2022corrupted}. To address this, researchers need to develop and adopt advanced filtering techniques capable of removing this periodic components before applying fractal-based approaches~\cite{nagarajan2005minimizing}. Only such detrending can ensure that the observed scaling accurately reflects the intrinsic dynamics of the system. 

Furthermore, given the fact that many air traffic data sets exhibit heavy-tailed distributions rather than Gaussian ones (as discussed in Sec. \ref{sec2}), it is important to recognise how this properties impact in the characterisation of the scaling behaviour. Heavy tails, i.e. higher probabilities of extreme values, can lead to systematic over- or under-estimations of the Hurst exponent, depending on the methodology being used~\cite{barunik2010hurst}.

Addressing these issues remains a critical step toward a reliable estimation of the scaling properties of traffic data that naturally exhibit a superimposed oscillatory and stochastic dynamics.

\begin{center}
\begin{table}[h!]
\begin{tabular}{|c| p{3.5cm} |p{0.5cm} |p{2.2cm}|p{2.2cm} |p{1.5cm}|} 
 \hline
Year & Title & Ref. & Geographic scope & Temporal scope &Type of data \\ [0.5ex] 
 \hline\hline
  2018 & Nonlinear dynamic analysis of air traffic flow at different temporal scales: nonlinear analysis approach versus complex networks approach  & \cite{wang2018nonlinear} & Sanya controlled airspace, China & May 28th to June 3th, 2016 & en-route Vol. \\ 
 \hline
 2019 & Multifractal detrended fluctuation analysis on air traffic flow time series: A single airport case & \cite{zhang2019multifractal} & China & Summer season 2017 & Arr., dep. and total Vol. \\ 
 \hline
 2020 & Characteristic analysis of flight delayed time series & \cite{lan2020characteristic} & Unspecified & 2014-2018& Number, rate and average of Arr. and Dep. delays\\
 \hline
 2020 & Multiscale multifractal analysis on air traffic flow
time series: A single airport departure flight case & \cite{liu2020multiscaleB} & China & Summer season 2017 & Dep. vol. \\
 \hline
 2022 & Corrupted bifractal features in finite uncorrelated power-law distributed data & \cite{olivares2022corrupted} & Europe &May 2018 to Dec. 2019 & Time intervals between landing\\
 \hline
 2023 & Measuring landing independence and interactions using statistical physics & \cite{olivares2023measuring} & Europe/China & May 2018 to July 2021 & Time intervals between landing \\ [1ex] 
 \hline
\end{tabular}
\caption{List of all papers using fractal and multi-fractal analysis for characterising air traffic data.}
\label{tab:MF}
\end{table}
\end{center}

\section{Nonlinear dynamics tools}\label{NLDT}

Air traffic networks are intricate dynamical systems where delays ripple through space and time in ways that are often non-linear \footnote{In mathematical terms, a function $f$ (in the case of this example, the delays observed at one airport given the delays in a second one) is defined as linear if it satisfies $f(\alpha x + \beta y) = \alpha f(x) + \beta f(y)$. In more intuitive terms, the function is not linear when the change in the output is not proportional to the change in the input.} and difficult to predict. From the cascading effects of a single weather event to the subtle interplay between airport efficiency and network congestion, the propagation of delays exhibits characteristics that traditional linear models may struggle to capture~\cite{fleurquin2013systemic}. In recent years, tools from nonlinear dynamics, especially the largest Lyapunov exponent~\cite{wolf1985determining} and the correlation dimension~\cite{grassberger1983characterization}, have emerged as powerful methods for a macro-scale analysis of aggregated time series representing the dynamics of such systems. These measures, rooted in chaos and nonlinear dynamics theory, quantify the unpredictability and geometric complexity.

\subsection{The largest Lyapunov exponent and correlation dimension}

The largest Lyapunov exponent (LLE) quantifies the sensitivity to initial conditions, by measuring how quickly nearby trajectories in the phase space diverge~\cite{kantz2003nonlinear} \footnote{How many Lyapunov exponents are there? Any $n$-dimensional system will have $n$ Lyapunov exponents, i.e. one for each dimension. For the sake of simplicity, the number of dimensions can be understood as the number of time series analysed at the same time. To illustrate, if one analyses a single time series, the result can only be one exponent, which, by definition, will also be the largest one (i.e. the LLE). On the other hand, let us suppose the simultaneous (multivariate) analysis of the departure and arrival time series of an airport; the dimensionality of this small system will be two, hence two Lyapunov exponents will be extracted. Most importantly, the sensitivity to initial conditions of this system only depends on the largest of the two, hence the importance of the LLE.}. In the context of air traffic, this can be interpreted as how small changes, such as a minor delay at one hub, can amplify and propagate across the network. A positive Lyapunov exponent is a necessary condition for chaotic behaviour; even tiny differences in flight schedules can result in a significantly different traffic flow. If, instead, the LLE is zero, the system exhibits a quasi-periodic behaviour characterised by the interaction of multiple incommensurable frequencies, i.e., the existence of patterns that are complex but still governed by predictable cycles, such as daily peaks in air traffic. Finally, a negative exponent means stability. 

Complementing the Lyapunov exponent is the correlation dimension ($D_2$), which quantifies the geometric complexity of the phase space structure shaped by the long-term evolution~\cite{kantz2003nonlinear}. This reveals how many degrees of freedom (independent variables) effectively drive the system's dynamics - to illustrate, how many variables are required to describe an airport's dynamics. For chaotic systems, one finds a non-integer dimension, which suggests a hidden structure underneath the apparent randomness. In general terms, this means that the system is not unpredictable but rather low-dimensional chaotic, with a few interacting variables (e.g., in the case of air transport, weather, congestion, or technical issues). 

The method by Wolf {\it et al.}~\cite{wolf1985determining}, based on phase space reconstruction and trajectory divergence, was the first practical estimator for LLE. Later, Rosenstein and co-workers~\cite{rosenstein1993practical} proposed a simplified approach better suited for short and noisy data sets, while Kantz~\cite{kantz1994robust} improved the robustness by using local neighbourhood divergence. On the other hand, The Grassberger–Procaccia algorithm~\cite{grassberger1983characterization} estimates the correlation dimension by evaluating how the number of near point pairs scales with distance in the reconstructed phase space. To correct for temporal correlations that may bias the results, Theiler~\cite{theiler1986spurious} introduced a windowing method that excludes temporally adjacent points.

\subsection{Applications}

As far as can be determined from existing studies, the estimation of the LLE and $D_2$ has only been applied to traffic flow volume data sets. Both en-route and arrival volumes exhibited a positive LLE, indicating a signature of chaos~\cite{yang2023chaotic,zhang2020data,zhang2024research}; in practical terms, this suggests that short-term forecasting remains feasible, while long-term predictions may be unreliable. Moreover, Zhang {\it et al.}~\cite{zhang2020data,zhang2024research} found that the degree of chaos intensifies at coarser time resolutions (2,5,10 and 15 minutes), meaning that a larger sampling window captures more information about the underlying chaotic nature. Wang {\it et al.}~\cite{wang2018nonlinear} not only confirmed this finding but also estimated the correlation dimension of en-route traffic volumes, finding a non-integer $D_2$, which validates a signature of chaos. Conversely, Cong and Hu~\cite{Cong2014chaotic} showed that for en-route traffic volume sampled at one minute, the LLE is equal to zero, yet the correlation dimension is non-integer. While these findings contradict the above results, the null LLE can be an artefact of the small temporal resolution used. Furthermore, the dynamics observed at this time scale can be interpreted as the air traffic flows generally operating in an orderly manner under the guidance of controllers, with occasional unpredictable perturbations. 

\subsection{Lessons learned and future directions}

Estimating the LLE from experimental time series represents a challenging task due to several limitation of the real-world data. Unlike numerical simulations, experimental time series suffer from noise contamination, finite length and low resolution, which difficult the accurate reconstruction of the real system's dynamics~\cite{wolf1985determining}. All these could lead to a mistaken estimation of the LLE, and therefore to a wrong classification of the underlying dynamical properties. Alternative methods, such as permutation entropy, offer practical solutions for analysing chaoticity in one-dimensional time series from experimental and man-made systems, when traditional methods are hindered by data limitations.

\begin{center}
\begin{table}[h!]
\begin{tabular}{|c| p{3.5cm} |p{0.5cm} |p{2.2cm}|p{2.2cm} |p{1.5cm}|} 
 \hline
Year & Title & Ref. & Geographic scope & Temporal scope &Type of data \\ [0.5ex] 
 \hline\hline
  2014 & Chaotic characteristic analysis of air traffic system & \cite{Cong2014chaotic} & Guangzhou area, China & 14th-18th Nov. no year specified  & en-route Vol. \\ [1ex] 
 \hline
 2018 & Nonlinear dynamic analysis of air traffic flow at different temporal scales: nonlinear analysis approach versus complex networks approach  & \cite{wang2018nonlinear}  & Sanya controlled airspace, China & May 28th to June 3th, 2016& en-route Vol. \\ 
 \hline
  2020 & Data-Driven Analysis of the Chaotic Characteristics of Air Traffic Flow & \cite{zhang2020data} & Laiyang city area, China & December 14th 2018 & en-route Vol. \\ 
 \hline
   2023 & A Chaotic Discriminant Algorithm for Arrival Traffic Flow Time Series Based on Improved Alternative Data Method & \cite{yang2023chaotic}& O'hare airport, US & August 2019 & arrival Vol. \\ 
 \hline
    2024 & Research in Chaotic Characteristics and Short-term Prediction of en-route traffic flow using ADS-B data & \cite{zhang2024research} & Shanghai, China & December 9th-15th 2020 & en-route Vol. \\ 
 \hline
\end{tabular}
\caption{List of papers using the Larguest Lyapunov Exponent and correlation dimension for characterizing air traffic data.}
\label{tab:MF}
\end{table}
\end{center}

\section{Discussion and conclusions}\label{sec5}

Statistical physics is not only a discipline dealing with abstract concepts like ideal gases or thermodynamics; through the exploration of such topics, its practitioners have developed a large set of tools that are well-suited for the analysis of many complex real-world systems. From entropy to fractality, the underlying leitmotiv is the extraction of insights about the micro-scale (i.e. what the individual composing elements are doing) when only access to the macro-scale (i.e. the overall system) is available. As seen throughout this review, statistical physics can and has been used in air transport, specifically with the aim of extracting knowledge from coarse-grained time series representing different aspects of operations, from departures and landings, to delays.

\begin{figure}[!tb]
\includegraphics[width=\textwidth]{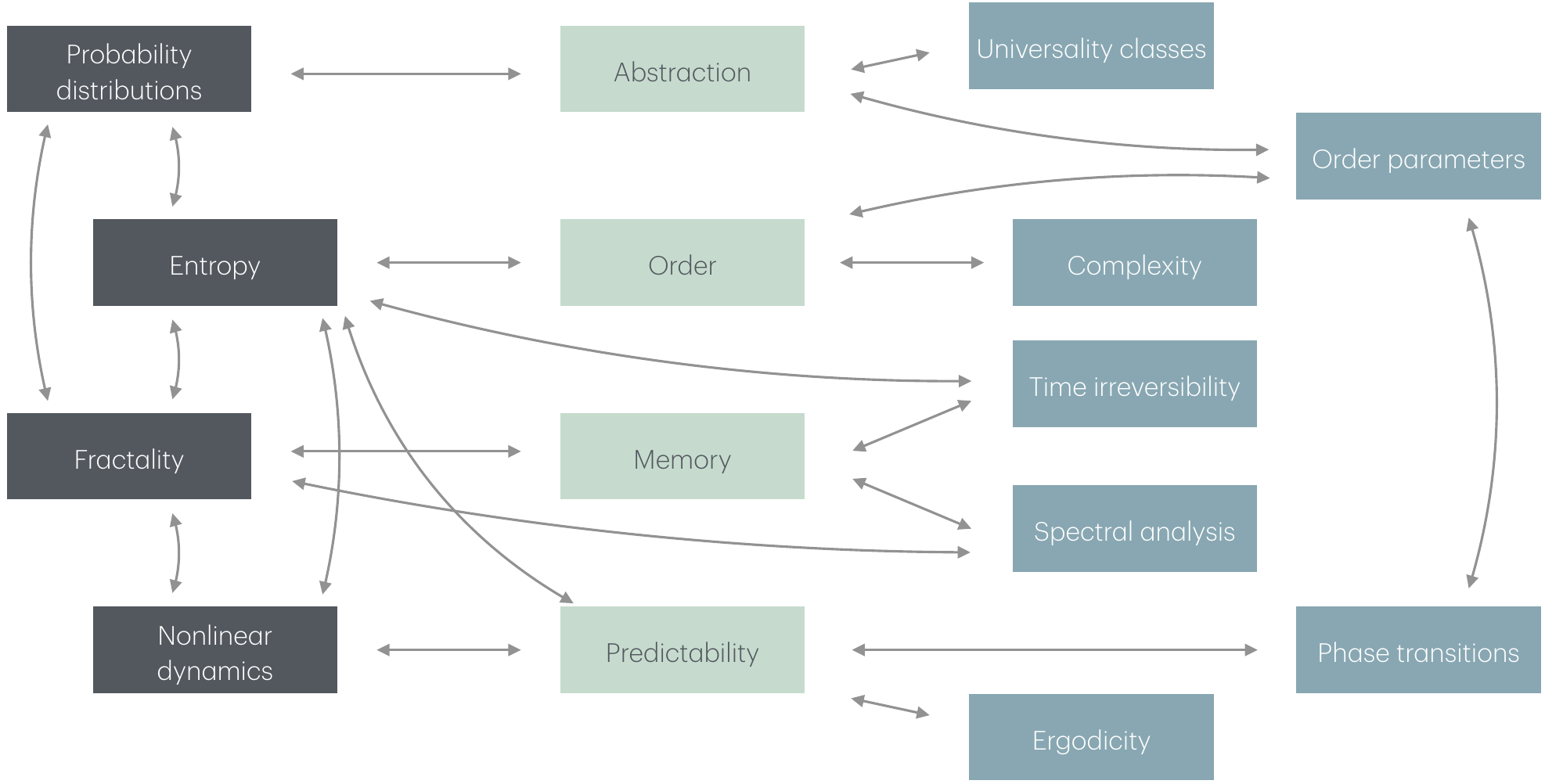}
\caption{ Concepts and their mutual relationships. Boxes represent the main concepts described in this review (left), what they represent from a more abstract viewpoint (centre), and the additional topics identified in Sec. \ref{sec:future_topics} (right). Arrows connect pairs of concepts that share objectives, characteristics or requirements. Note that this represents a simplified view, and should only be used for illustration purposes.
\label{fig:FinalDiagram}}
\end{figure}

If several notable results have here been reported, the attentive reader would also have noted many challenges. In order to conclude this work, we are here going to discuss four points that we believe should be taken into account by any practitioner:

\begin{enumerate}

    \item The metrics and concepts here described are not independent, but rather form an intricate mesh of interconnections - i.e. a complex network, another foundational concept in statistical physics \cite{strogatz2001exploring}. Entropies can yield information about the dynamics of a system in terms of its predictability and regularity; yet, its chaotic nature has to be evaluated through other non-linear tools, like the Lyapunov exponent and the correlation dimension. Metrics based on fractality can elucidate the presence of long-range memories, but these may be biased both by heavy-tailed distributed data and by regular oscillations, which have respectively to be assessed through probability distributions and multiscale entropies. In addition, connections with other scientific fields are beneficial: to illustrate, long-range memories impact the fractality of the system, but can also be measured through tools from Information Theory, as e.g. the Active Information Storage \cite{lizier2012local}. In short: while most works here reviewed have analysed elements of the air transport system using individual metrics, a more holistic view is needed, in which multiple metrics are combined to answer specific questions.

    \item It may at first be surprising that rather different results have been obtained for the same question, even when using relatively simple tools. The clearest example of this is the characterisation of delays using probability distributions: as seen in Sec. \ref{sec2}, results spanned from Gaussian \cite{mueller2002analysis, tu2008estimating, wang2022distribution}, to Poisson \cite{mueller2002analysis}, exponentials \cite{mitsokapas2021statistical}, Student's t \cite{szabonon}, and power laws \cite{wang2020universal, mitsokapas2021statistical}. Regardless of the cause behind this (i.e. data heterogeneity or technical considerations), the air transport community has to start tackling this issue. To illustrate, instead of analysing any delay data available, practitioners have to question how these delays have been calculated, and check whether other definitions yield consistent results. This will be essential both towards reproducibility and trustworthiness.

    \item There is a clear gap between the results obtained in the reviewed works, and the operational aspects of air transport. In other words, while some works have reported concrete results, e.g. related to the efficiency of individual airports, translating these to operational improvements is a task far from simple. This is of course not a problem specific to air transport: knowing that a virus causes a disease is (unfortunately) not tantamount to have a cure for it. Additionally, one must not underestimate the importance of obtaining new knowledge about the system, even though its impact may be far in the future. Still, a balance ought to be achieved, possibly by integrating the viewpoint of operational experts in these theoretical analyses.
    
    \item The concepts and metrics here explored are not the only available ones, as they have been selected taking into account their usage in the literature. Statistical physics provides many other points to start new discussions, and to shed light on air transport from different angles. In what follows we briefly discuss what we see as the potential way ahead. Note that, following the first point of this list, these are not independent concepts: they have instead to be integrated with what here discussed, and with other views to the air transport system, to yield a more complete picture. For the sake of illustration, a simplified representation of these relationships is sketched in Fig. \ref{fig:FinalDiagram}.
    
\end{enumerate}

\subsection{The way ahead: future topics and concepts}\label{sec:future_topics}

\vspace{0.5cm}
\noindent
{\bf Entropy {\it vs.} complexity.} As already discussed in Sec. \ref{sec3}, and in spite of the large number of works assuming such relationship, entropy does not necessarily equate to complexity. While the former quantifies the degree of disorder in a system, the latter focuses on its degree of organisation. To illustrate the difference, suppose a system displaying a periodic dynamics: it would be very easy to predict (hence low entropy), and will further have a trivial temporal structure (hence low complexity); on the other hand, a system exhibiting a random behaviour would be impossible to predict (high entropy), but at the same time, it would show no structure (also low complexity). Statistical physics provides several tools to estimate the complexity of a system - even though it can also be acknowledged that any definition of this term is somewhat incomplete. These range from metric combining the information encoded in the system and its disequilibrium \cite{lopez1995statistical, feldman1998measures}; the amount of information about the past required to predict the future (and hence, the intrinsic computation of the system) \cite{crutchfield1989inferring}; or the size of the smallest algorithm able to mimic the system \cite{kolmogorov1965three, chaitin1966length}.

\vspace{0.5cm}
\noindent
{\bf Spectral analysis.} When thinking about the dynamics of an element or of a system, the most natural representation that comes to mind is a time-domain signal, i.e. the evolution of a metric through time. The complementary frequency-domain representation, as e.g. obtained through a Fourier Transform, can nevertheless provide invaluable information. In spite of its relevance, not least in neighboring fields like material science, the use of spectral analysis in air transport has been limited to a handful of works \cite{ge2020measuring, diana2021doing, barczak2022impact}. Spectral analysis describes how the power is distributed over different frequencies; in the case of stochastic processes, such power can be understood as the variance. This can be used to understand the time-scales of correlations in the dynamics, and hence of the memory - as, according to the Wiener-Khinchin Theorem \cite{wiener1930generalized}, these are equivalent in stationary random processes. To illustrate, a higher power at low frequencies indicate that long-lasting correlations, and in general repetitive patterns, are present. When the system under analysis is composed of multiple elements connected together, oscillations (i.e. spectral peaks) are associated with the normal modes of vibration; additionally, when the system is driven by an external force, resonances can help in the identification of frequencies for which such forcing is more effective. Thinking on air transport and its delays, such resonances may indicate that specific disruptions may have an impact much larger than expected, as they resonate with the normal dynamics of the system, and have hence to especially be avoided. Finally, multiple interconnected systems may share information (i.e. may be coupled) at only specific frequencies, which can be detected using cross-spectral analyses \cite{vowels2023spectral}. To illustrate, delays may propagate between airports at high frequencies only, i.e. over short time scales and in bursts; such propagation may be lost when analysing the time-domain signal, and yet clearly appear when using frequency-domain metrics - as e.g. the Partial Directed Coherence (PDC) \cite{baccala2001partial} or the Directed Transfer Function (DTF) \cite{kaminski1991new}.

\vspace{0.5cm}
\noindent
{\bf Phase transitions.} An essential concept in complex systems and beyond is that of phase transitions, i.e. when the smooth change in a control parameter triggers a sudden and abrupt change in the system itself \cite{landau1936theory, sole2011phase}. One may thing, for instance, on the classical example of water being cooled down: after passing zero degrees, there is a transition between liquid and solid phases, where the properties of the water/ice suddenly and substantially change. The researcher experienced in air transport will surely have found similar events: a day initially normal, in which suddenly delays start propagating and snowballing, without a clear external trigger event.
Statistical physics yields many tools and concepts to understand such phenomena, from the identification of critical points, i.e. combinations of parameters where the transition takes place; to the description of symmetries inside the system, and how these are broken during a transition \cite{strocchi2005symmetry}.

\vspace{0.5cm}
\noindent
{\bf Order parameters.} A concept connected to the previous idea of phase transitions is that of order parameters, i.e. some metrics (or observable) of the system that is able to describe its current phase. Consider the previous example of water turning into ice: analysing the density will tell us if we are looking at the liquid or the solid phase. Delays can also be described through order parameters, for instance by calculating the relationship between primary and reactionary delays: this metric can distinguish between a phase in which random events generate random and uncorrelated disruptions, and one in which the system cannot cope with them and is dominated by increasing propagations.

\vspace{0.5cm}
\noindent
{\bf Time irreversibility.} 
In classical thermodynamics, the arrow of time paradox establishes that, while the microscopic laws of motion are time-reversible, macroscopic processes, such as heat flow or diffusion, proceed in one preferred direction~\cite{parrondo2009entropy}. At the level of individual molecules, e.g. gas particles obeying Newton's equations, the dynamics are perfectly symmetric under time reversal: if you reversed every velocity, the system would retrace its steps exactly. Yet in practice, we never see a cold cup of coffee rewarm itself by drawing in heat from the cooler air, or a chamber of mixed gases spontaneously unmix~\cite{lebowitz1993}. Translating this idea to a single time series, we say the evolution is temporally irreversible if the statistical patterns found forward in time do not match when observing the same data backward~\cite{weiss1975time}; in other words, the time asymmetry is broken. In air traffic data, this temporal asymmetry is easy to spot. Imagine a day when delays emerge after a morning storm, to then propagate and slowly dissipate; the hourly arrival count would abruptly decrease for then gradually recovering, hence breaking the forward/backward symmetry. Irreversibility in the time series excludes Gaussian linear dynamics as a valid model, and in fact, it embraces the presence of nonlinearities in the underlying dynamics~\cite{lacasa2012time}. When a system is constantly driven by external inputs out of equilibrium~\cite{gnesotto2018broken}, such as bad weather forcing to cancel flights, and dissipates energy or “traffic pressure” unevenly, we face a break in time‐reversal symmetry of the driving and relaxation processes~\cite{kawai2007dissipation}.

Recognising temporal irreversibility in air traffic time series is more than a curiosity. It reveals the directional memory of the system, the one-way streets of congestion and recovery that models must capture in order to predict delays. Yet, measuring irreversibility in real data is challenging, both theoretically and numerically. Note that the previous definition referred to ``statistical patterns'', and indeed imposes no restriction on what these patterns may be; consequently, many complementary (and at times, contradicting) tests have been proposed in the last decades \cite{daw2000symbolic,zanin2018assessing,brock1991nonlinear,broock1996test,lacasa2012time,costa2005broken,kennel2004testing,diks1995reversibility,gaspard2004time,cammarota2007time,martinez2018detection} - the interested reader can found reviews, comparisons, and software implementations in Refs. \cite{zanin2021algorithmic, zanin2025algorithmic}.
To the best of our knowledge, only three works have used irreversibility tests in the context of air transport. Using two different metrics, Refs. \cite{martinez2023complementarity} and \cite{zanin2021assessing} respectively detected low and large irreversibility on delay time series, with the former observing higher asymmetry for small airports. Although partly contradictory, these findings point to the presence of delay propagation mechanisms and memory, and this challenges the notion of flight delays as purely stochastic events.
In a recent analysis~\cite{olivares2023measuring}, the authors investigated the irreversibility of inter-landing time sequences to diagnose interaction dynamics in airport operations at 12 European and 10 Chinese airports. The study showed that irreversibility emerges as an indicator of memory in landing time intervals, indicating a deviation from purely independent operations (null model). Comparisons made between the descent and landing phases revealed that most irreversibility emerges close to the runway, linking it to procedural or sequencing constraints.

\vspace{0.5cm}
\noindent
{\bf Universality classes.} A universality class can be seen as a collection of systems, or of mathematical models thereof, that are radically different when analysed at a micro-scale; but that nevertheless give rise to the same macro-scale properties. To illustrate, consider the evolution of delays at a given airport; even though each day may have a different planning, and random events alter such planning at a micro-scale in ever different ways, the final macro-scale evolution of the delays is quite consistent \cite{ivanoska2022assessing, cuerno2025topological}. The identification of universality classes can help pushing the analysis in two opposite directions. On the one hand, it simplifies the understanding of the system: the micro-scale details can be eliminated, to only focus on the macro-scale. Back to previous example, if the global evolution of delays is a universality class, these can be studied independently of the specific events that originated them. On the other hand, it is also interesting to study the scale at which the universality class emerges - in other words, one may ask what zoom has to be applied to the system, before delays stop evolving according to universal patterns.

\vspace{0.5cm}
\noindent
{\bf Ergodicity.} At the foundation of statistical physics, ergodicity describes the properties of systems for which the time and ensemble average of a given quantity are equivalent \cite{hawkins2021ergodic}. Imagine analysing a property of air transport, as e.g. the number of aircraft operating within a time window. The time average refers to the estimation of this number over a long time window, e.g. throughout one year. On the other hand, the ensemble average can be estimated by calculating the same metric in a large number of short windows, e.g. of one day, and average the corresponding results. Whenever these two averages are not equal, the ergodicity is broken \cite{palmer1982broken}, with several mechanisms being potential causes: the path-dependence of the system, i.e. the dependence of its dynamics on the past history; or the presence of multiple stable states. Ergodicity breaking implies that forecast models may not be reliable, especially if based on averages of past data. Additionally, such lack of ergodicity may be localised at specific spatial scales, thus connecting back to the previous concept of spectral analysis.

\section*{Acknowledgements}

This project has received funding from the European Research Council (ERC) under the European Union's Horizon 2020 research and innovation programme (grant agreement No 851255). This work was partially supported by the Mar\'ia de Maeztu project CEX2021-001164-M funded by the  MICIU/AEI/10.13039/501100011033 and FEDER, EU. This project has received funding from Grant CNS2023-144775 funded by MICIU/AEI/10.13039/501100011033 by ``European Union NextGenerationEU/PRTR''.

\bibliography{sn-bibliography}

\end{document}